\documentclass[review,5p,12pt,twocolumn]{elsarticle}

\usepackage{amssymb}
\usepackage{amsmath}
\usepackage{graphicx}
\usepackage{color}

\begin{document}

{\begin{frontmatter}

%% Title, authors and addresses

%% use the tnoteref command within \title for footnotes;
%% use the tnotetext command for theassociated footnote;
%% use the fnref command within \author or \address for footnotes;
%% use the fntext command for theassociated footnote;
%% use the corref command within \author for corresponding author footnotes;
%% use the cortext command for theassociated footnote;
%% use the ead command for the email address,
%% and the form \ead[url] for the home page:
%% \title{Title\tnoteref{label1}}
%% \tnotetext[label1]{}
%% \author{Name\corref{cor1}\fnref{label2}}
%% \ead{email address}
%% \ead[url]{home page}
%% \fntext[label2]{}
%% \cortext[cor1]{}
%% \address{Address\fnref{label3}}
%% \fntext[label3]{}

\title {Reaction pathway of oxygen evolution on Pt(111) revealed through constant Fermi level molecular dynamics}

%% use optional labels to link authors explicitly to addresses:
%% \author[label1,label2]{}
%% \address[label1]{}
%% \address[label2]{}

\author{Assil Bouzid\corref{mycorrespondingauthor}}
\cortext[mycorrespondingauthor]{Corresponding author}
\ead{assil.bouzid@unilim.fr}
\author{Patrick Gono}
\author{Alfredo Pasquarello}
\address{Chaire de Simulation \`a l'Echelle Atomique (CSEA), Ecole Polytechnique F\'ed\'erale de Lausanne (EPFL), CH-1015 Lausanne, Switzerland}

\begin{abstract}
The pathway of the oxygen evolution reaction at the Pt(111)/water interface is disclosed through constant Fermi level molecular dynamics. %20
Upon the application of a positive bias potential ${\rm H_2O}_{\rm ads}$ and ${\rm OH}_{\rm ads}$ adsorbates are found to arrange
in a hexagonal lattice with an irregular alternation. %24
Increasing further the electrode potential then induces the oxygen evolution reaction, which is found to proceed through a hydrogen
peroxide intermediate. %21
Calculation of the associated overpotential shows a reduction of 0.2 eV compared to the associative mechanism. %16
This result highlights the forcefullness of the applied scheme in exploring catalytic reactions in an unbiased way. %17
%Count: 20+24+21+16+17= 98
\end{abstract}

\begin{keyword}
%% keywords here, in the form: keyword \sep keyword

%% PACS codes here, in the form: \PACS code \sep code

%% MSC codes here, in the form: \MSC code \sep code
%% or \MSC[2008] code \sep code (2000 is the default)
OER mechanism\sep molecular dynamics\sep constant Fermi level molecular dynamics
\end{keyword}

\newpage
\end{frontmatter}

%% \linenumbers

%% main text
\newpage

%Fuel cells and Pt electrode
Reducing CO$_2$ emissions is becoming a worldwide necessity in order to limit the impact of global warming.
This requires developing new energy sources that go beyond internal combustion engines \cite{fontaras2007quantitative,basha2009review,samanta2011post}.
Alkaline and proton-exchange-membrane fuel cells are among the most promising alternatives thanks to their
high energy efficiency \cite{fontaras2007quantitative,basha2009review,samanta2011post}. Enhancing the performance of fuel cells requires a catalyst 
that can efficiently catalyze the hydrogen and oxygen evolution at the same time. While platinum remains the best electrocatalyst for fuel cells \cite{kinoshita1992electrochemical},
many efforts focus on developing efficient and low-cost electrode materials. At this level, a thorough understanding of the oxygen evolution reaction (OER) mechanism 
under realistic conditions is instrumental for further progress.

%OER mechnism : associative
The reverse reaction, the oxygen reduction reaction, is generally assumed to evolve through four proton-coupled electrons transfers
according to an associative mechanism \cite{norskov2004origin}. The pathway of the OER can then be inferred under the assumption that 
the same intermediates occur. In alcaline conditions, this gives: 
\begin{align*}
^{*} + {\rm H_2O}              &\rightarrow {\rm OH}_{\rm ads} + ({\rm H}^{+} + e^-), \\
{\rm OH}_{\rm ads}             &\rightarrow {\rm O}_{\rm ads} + ({\rm H}^{+} + e^-),  \\
{\rm O}_{\rm ads} + {\rm H_2O} &\rightarrow {\rm OOH}_{\rm ads} + ({\rm H}^{+} + e^-),  \\
{\rm OOH}_{\rm ads}            &\rightarrow {\rm O}_{2\rm g} + ({\rm H}^{+} + e^-), 
\end{align*}
where the asterisk indicates an active surface site. First, water molecules dissociate leading to hydroxyl ions adsorbed at the metal surface and
hydronium ions in the aqueous solution. Second, ${\rm OH}_{\rm ads}$ releases a proton and forms an adsorbed O. Third, ${\rm OOH}_{\rm ads}$
forms upon the next proton-coupled electron transfer. Finally, an oxygen molecule is formed and released from the metal surface.
The high stability of the intermediate step with ${\rm O}_{\rm ads}$ is assumed to determine the overpotential of the overall reaction \cite{norskov2004origin}.
%OER mechnism: dissociative
%At variance, the dissociative reaction differs in the third and fourth steps involving the following intermediates:
%\begin{align*}
%{\rm O}_{\rm ads} + {\rm H_2O} &\rightarrow {\rm O}_{\rm ads} + {\rm OH}_{\rm ads} + ({\rm H}^{+} + e^-),  \\
%{\rm O}_{\rm ads} + {\rm OH}_{\rm ads}            &\rightarrow {\rm O}_{2\rm g} + ({\rm H}^{+} + e^-), 
%\end{align*}
%In this case, ${\rm OOH}_{\rm ads}$ is formed from two ${\rm OH}_{\rm ads}$ upon a proton-coupled electron transfer.\cite{ramaswamy2012fundamental,appleby1993electrocatalysis}

%role of adsorbates
In the OER pathway given above, the detailed role of intermediate adsorbates, such as ${\rm OH}_{\rm ads}$, remains elusive. 
On the one hand, it has been shown that ${\rm OH}_{\rm ads}$ slows down the kinetics of the OER by  
blocking the catalytically active sites \cite{li2013ph,roques2005potential}. On the other hand, it facilitates the reaction by 
stabilizing ${\rm H_2O}_{\rm ads}$ through the formation of hydrogen bonds \cite{bondarenko2011pt,karlberg2005interaction,liu2016mechanism}. 
The reaction could further complexify through the formation of a hydrogen peroxide (${\rm H_2O}_{2\rm ads}$) intermediate as proposed 
by Wroblowa \textit{et al}.~\cite{wroblowa1976electroreduction} and supported by several experimental 
observations \cite{gomez2012interaction,zhang2005controlling,stamenkovic2001structure,schmidt2001oxygen,schneider2008transport,seidel2009mesoscopic,markovic2002surface,markovic1999oxygen,grgur1997temperature}.

%Modelling efforts 
While assessing the OER mechanism from experiments remains difficult \cite{wang2004kinetic}, density functional theory (DFT) can provide valuable and detailed
insights into the reaction pathways and their energetics \cite{norskov2004origin,tripkovic2010oxygen,liu2016mechanism,tripkovic2011standard,skulason2007density,rossmeisl2013ph}.
However, current modeling techniques in which the number of electrons remains constant do not provide direct access to the reaction mechanisms, and rely on a prior knowledge of
the reaction intermediates. It is therefore highly desirable to implement modeling schemes that allow chemical reactions to take place spontaneously
at finite temperature and under bias potential \cite{bouzidjctc2017,bouzid2018atomic,otani2006first,otani2008structure,otani2008electrode,ikeshoji2017first}.
Such techniques would make possible the exploration of intermediate species and the investigation of their role in the OER as a function 
of the applied potential in a dynamically unconstrained fashion \cite{bouzid2018atomic}. 

%In this Letter
In this Communication, we explore the use of constant Fermi level molecular dynamics \cite{bonnet2012first,bouzid2018atomic} 
to investigate electron-transfer reactions at metal-water interfaces under bias potential, focusing in particular on the 
OER at Pt(111) electrodes. We pursue the following two-step strategy. First, we use constant Fermi 
level molecular dynamics to allow the system to evolve spontaneously without taking advantage of any 
previous knowledge that could bias the outcome of the simulation. Accordingly, we study the Pt surface 
coverage under positive bias potential and discuss the spontaneous arrangement of the adsorbates. By further 
increasing the applied potential, we then induce the OER at the Pt surface and explore the pathway freely undertaken 
by the system. Next, we determine the free energies of the identified intermediates within a conventional scheme, 
namely the computational standard hydrogen electrode, in order to examine whether the revealed pathway is viable. 

%\textcolor{red}{We note that, from a modelling perspective the intermediate steps of the OER reaction are the same as 
%those of the oxygen reduction reaction (ORR). In the present work we focus on the OER reaction as it does not require 
%the knowledge of the Pt surface properties at the atomic scale, which will be spontaneously formed during the simulation.}

\begin{figure}[!t]
\begin{center}
\includegraphics*[width=1.0\linewidth, keepaspectratio=true]{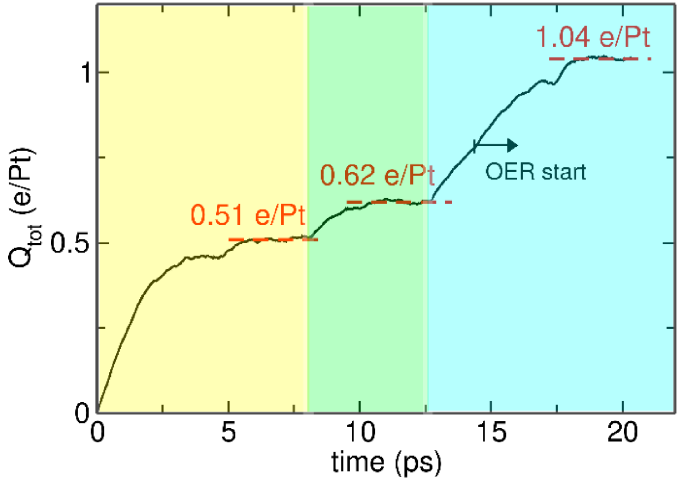}
\caption{Time evolution of the excess electronic charge $Q_{\rm tot}$ in constant Fermi level molecular
dynamics of the Pt(111)/water interface. $Q_{\rm tot}$ is given per number of Pt atoms.
The background colors indicate three different potential regimes.}
\label{Fig1}
\end{center}
\end{figure}

%Constant-Fermi-Level MD 
In constant Fermi level molecular dynamics, the Fermi level is maintained at a preset fixed value 
during the evolution and is therefore particularly suited for simulating the effect of the bias 
potential in the simulation of electrified metal-water interfaces \cite{bonnet2012first,bouzid2018atomic}. 
This is achieved through the introduction of grand canonical potential, by which the 
charge can flow in and out of the system so that the condition on the Fermi level is satisfied. 
In practice, the electronic charge in the system is treated like a classical variable with inertia 
and is subject to fictitious dynamical equations. To ensure a good control over the potential fluctuations, 
it is also possible to couple the physical system to a thermostat ensuring a tempurature $T$, while 
keeping the electron reservoir at a temperature $T_e$ \cite{bonnet2012first}.
The use of two different thermostats does not create any significant thermal flow in the system
due to the weak coupling between atoms and charges \cite{bonnet2012first}.
When applying constant Fermi level simulations to reactions occurring at metal-water 
interfaces, it should be kept in mind that this corresponds to the modeling of a rare event with
a computational setup of relatively small size and on time scales that can be reached with molecular 
dynamics techniques. In the present approach, the intrinsic barriers associated to these events are 
overcome by increasing the bias potential until the reaction is seen to proceed spontaneously.
While enforcing the reaction in this way appears sufficient to reveal interesting reaction pathways in this work,
it could be envisaged to combine the present constant Fermi level molecular dynamics with more sophisticated 
approaches for treating rare events, such as metadynamics, Blue Moon, or umbrella sampling techniques.

%Our model
We adopt the periodic Pt(111)/water interface model generated in Ref.\ \citenum{bouzid2018atomic}, in which
a Pt slab shows two equivalent interfaces with the water layer, denoted surface 1 and surface 2. 
The supercell contains three Pt layers of 12 atoms with a $3\times \sqrt{3}$ surface repeat unit and 32 water molecules,
corresponding to a size of 9.79$\times$8.48$\times$19.50 {\AA}$^3$.
While the finite size of the water layer containing 32 water molecules only approximately 
accounts for the properties of liquid water at the interface, the O-O pair correlation function and
the water density in the middle of the slab are in fair agreement with the corresponding properties 
of bulk water \cite{bouzid2018atomic}. Furthermore, the dipoles of the water molecules remain randomly 
oriented even under bias potential. Therefore, in view of the fact that we focus in this work on electrochemical 
processes occurring at the metal-water interface, we assume that the finite size of the water layer only 
induces a limited bias in identifying the relevant reaction pathways. Furthermore, the adopted simulation cell
combined with the use of a uniform compensating background charge yields the advantage that the electric field 
is realistically modelled at the Pt/water interface ({\it vide infra}) \cite{bouzid2018atomic}.
In any event, we remark that our strategy comprises an a posteriori validation of the identified intermediate 
steps within a conventional state-of-the-art scheme.

%This corresponds to the minimal size required to model interfacial reactions in the presence of a central water layer
%with properties close to those of bulk water \cite{bouzid2018atomic}. It is here important to remark that this 
%is sufficient in the present study as the primary purpose of the constant Fermi level molecular dynamics is to 
%reveal reaction pathways in an unbiased fashion, whereas the ultimate energetic validation of the identified 
%intermediates would be separately analyzed a posteriori. 

%Electronic structure
The constant Fermi level molecular dynamics \cite{bouzid2018atomic,bouzid2017electron,bouzid2017identification,bouzidjctc2017} 
in this work relies on a description of the electronic structure within density functional theory (DFT) through an 
optimized rVV10 functional \cite{vydrov2010nonlocal,sabatini2013nonlocal,miceli2015isobaric}. 
In the rVV10 functional used in this work, the $b$ parameter is set to 9.3 to ensure a good description of 
the water density \cite{miceli2015isobaric}.
Core-valence interactions are treated through normconserving pseudopotentials \cite{TM} and
the wave functions of the valence electrons are expanded on a plane-wave basis set defined
by a kinetic-energy cutoff of 80 Ry. The Brillouin zone is sampled at the $\Gamma$ point.
Constant Fermi level molecular dynamics simulations at $T=350$ K are performed within the Born-Oppenheimer scheme with a
time step of $\Delta t = 0.48$~fs to integrate the equations of motion, as implemented in the {\sc q}uantum-{\sc espresso} suite of
programs \cite{giannozzi2009quantum}. Temperature control is ensured through velocity rescaling.
In the constant Fermi level scheme, the mass of the fictitious charge degrees of freedom
is set to $M_{\rm e}=100$ a.u. The fictitious electronic temperature is controlled by a velocity rescaling method
and is set to $T_{\rm e}=30$ K. Gaussian-smeared occupations with widths of at most 0.10 eV are used to prevent numerical
instabilities \cite{bouzid2018atomic,bouzid2017electron,bouzid2017identification,bouzidjctc2017}.
In the present dynamics, the inertial mass is set in such a way that the fluctuations of the grand canonical 
potential occur on a time scale that accompanies the structural rearrangements, to account for smooth transitions upon 
charge variations in the physical system. While this choice affects the time scales over which equilibrium 
is reached, the final state and its energetics are only determined by the grand-canonical potential.

When modeling a heterogeneous system consisting of an electrode-water system, a further issue arises concerning the
compensation of the electron charge on the electrode. 
Taylor et al.\ proposed a scheme based on a uniform background charge to 
compensate the electron charge injected in the system \cite{taylor2006first}. Applying an alignment procedure through 
a vacuum layer placed in the middle of the water layer allowed them to reference the electrode potential to the vacuum level.
In our approach, we also resort to a uniform background to ensure the overall neutrality of the simulation cell. 
However, we correct the charge and electrode potential through classical electrostatics and introduce an intrinsic 
alignment scheme based on the computational standard hydrogen electrode to reference 
the electrode potential \cite{bouzid2018atomic}.
It has been shown in Ref.\ \citenum{bouzid2018atomic}
that, for specific sizes of the simulation cell, the profile of the electric field across the interface closely 
resembles that of the physical interface in which the electrode charge is compensated by a distribution of counterions.
Hence, such an electrode model provides a convenient tool for investigating electrified 
metal-water interfaces.
%while awaiting affordable computational schemes for treating the couterions in an atomistic way.
More details about the computational setup and the model are provided in the Supporting 
Information of Ref.\ \citenum{bouzid2018atomic}.

We note that the use of a neutralizing background does not solve the general problem of modelling 
electrified interfaces from first principles. When modelling a half cell, one of the main difficulties consists 
in adapting the counter charge of the electrolyte as the electrode potential is varied. This implies inserting 
extra ionic charges into the electrolyte under conditions pertaining to a grand-canonical modelling scheme. 
When considering the results obtained with the more approximate scheme used in this work, it should therefore
be borne in mind that it is at present unknown to what extent the description of the alignments and of the reaction
mechanisms might be altered when treating the counterions in such an atomistic way.

Starting from a neutral system, we increment the bias potential towards positive values on the conventional scale
referred to the standard hydrogen electrode. In order to define the surface bias properly in our simulation it 
is necessary to determine the potential of the electrode with respect to the electrostatic potential in liquid water. 
However, as the reaction proceeds in the simulation, a large number of protons are injected in the water layer preventing 
us from recovering the potential of pure liquid water. Hereafter, we therefore characterize the Pt(111)/water 
system by the total electronic charge as this quantity is directly related to the applied potential \cite{bouzid2018atomic}.
To model increasing bias potentials, the total electronic charge goes through three plateaus as a function
of simulation time, being incremented by a discrete amount from one value to the next \cite{bouzid2018atomic}. Anytime the charge 
is varied, the system recovers equilibrium in a few picoseconds, as shown in Fig.\ \ref{Fig1}. This time period ensures 
that the system accommodates the injected electronic charge and that the water molecules arrange their structure accordingly.

\begin{figure}[!t]
\begin{center}
\includegraphics*[width=1.0\linewidth, keepaspectratio=true]{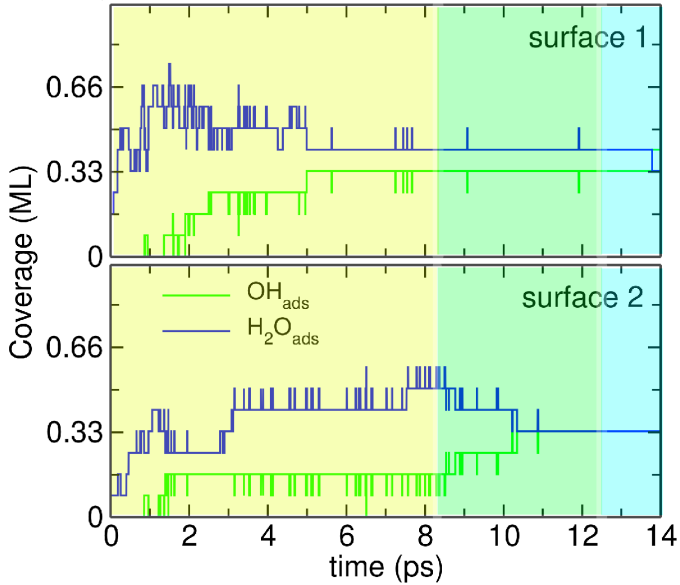}
\caption{Time evolution of the ${\rm H_2O}_{\rm ads}$ and ${\rm OH}_{\rm ads}$ surface coverage on 
the two individual surfaces in our model, surface 1 (top) and surface 2 (bottom).
The background colors indicate the three potential regimes.}
\label{Fig2}
\end{center}
\end{figure}

%surface coverage in general and OH coverage in particular
During the first part of the dynamics ($t<8$ ps) at a potential corresponding to 0.51 $e^-$/Pt (see Fig.\ \ref{Fig2}), 
part of the water molecules dissociate leading to the formation of adsorbed hydroxyl ions (${\rm OH}_{\rm ads}$) 
at the Pt surface and to the injection of hydronium ions (${\rm H_3O^{+}}$) into the water layer, 
as shown in Fig.\ \ref{Fig3}. We remark that despite the increase in acidity of the water layer, the surface 
coverage stabilized by the bias potential shows a high percentage of hydroxyls, typical of alcaline conditions.
As the electronic charge is increased further to 0.62 $e^-$/Pt ($8<t<12$ ps), 
the number of adsorbed species at the Pt surface reaches equilibrium. 
During the last part of the dynamics and before the OER takes place, only a few proton transfers between 
${\rm OH}_{\rm ads}$ and ${\rm H_2O}_{\rm ads}$ are observed at the Pt surface.
%
%During the dynamics, water molecules are either adsorbed at the Pt surface (${\rm H_2O}_{\rm ads}$) 
%or dissociated leading to adsorbed hydroxyl ions (${\rm OH}_{\rm ads}$)
%and aqueous hydronium ions (${\rm H_3O^{+}}$). As shown in Figs.\ \ref{Fig1} and \ref{Fig2}, 
%the Pt coverage reaches equilibrium when the potential is set at 0.62 $e^-$/Pt.
%
We find that the equilibrated surface structure corresponds to an ${\rm OH}_{\rm ads}$ coverage of 1/3 monolayer (ML). 
In addition, the ${\rm H_2O}_{\rm ads}$ coverage at equilibrium is found to be equal to 5/12 ML at surface 1 and 1/3 ML at surface 2. 
The slightly higher coverage of surface 1 is due to the adsorption of one extra water molecule at the Pt surface, 
indicated by an arrow in Fig.\ \ref{Fig3}. 
Overall, the ${\rm OH}_{\rm ads}$ coverage of 1/3 ML found through constant Fermi level molecular 
dynamics agrees with that predicted by static DFT calculations %
\cite{tripkovic2010oxygen,hansen2008surface,liu2016mechanism,bondarenko2011pt,karlberg2005interaction}. 
%thereby supporting the validity of our scheme. 

%surface structure and alternation of H2Oabs and OHads
The elusive role of adsorbates in the OER mechanism 
\cite{li2013ph,roques2005potential,bondarenko2011pt,karlberg2005interaction,liu2016mechanism} 
calls for a deeper understanding of their structural organization at the Pt active sites.
As shown in Fig.\ \ref{Fig3}, the adsorbates arrange in a honeycomb lattice, in agreement 
with previous predictions based on static calculations without bias potential and with a single monolayer
\cite{tripkovic2010oxygen,hansen2008surface,liu2016mechanism,bondarenko2011pt,karlberg2005interaction}.
However, ${\rm H_2O}_{\rm ads}$ and ${\rm OH}_{\rm ads}$ do not respect the perfect alternation as 
reported previously \cite{tripkovic2010oxygen,hansen2008surface,liu2016mechanism,bondarenko2011pt,karlberg2005interaction}.
Hence, it appears that irregular surface coverages can occur under the globally different conditions
of our simulation, which include the effects of finite temperature, bias potential and the interaction 
with water molecules beyond the first monolayer.

\begin{figure}[!t]
\begin{center}
\includegraphics*[width=1.0\linewidth, keepaspectratio=true]{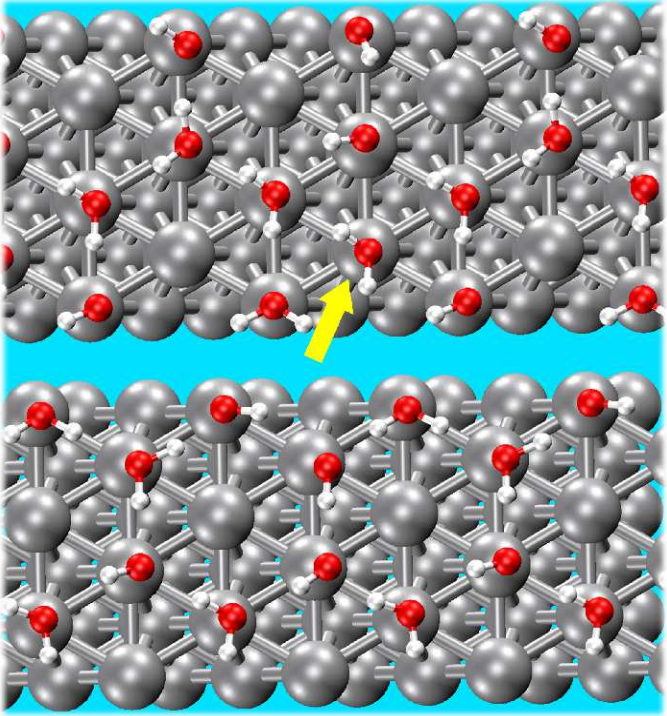}
\caption{Representative snapshots of Pt surface 1 (top) and Pt surface 2 (bottom) at equilibrium 
with $Q_{\rm tot}$=0.62 $e^-$/Pt (t=13 ps). The arrow indicates an extra water molecule adsorbed out of the hexagonal lattice.}
\label{Fig3}
\end{center}
\end{figure}

\begin{figure}[!t]
\begin{center}
\includegraphics*[width=1.0\linewidth, keepaspectratio=true]{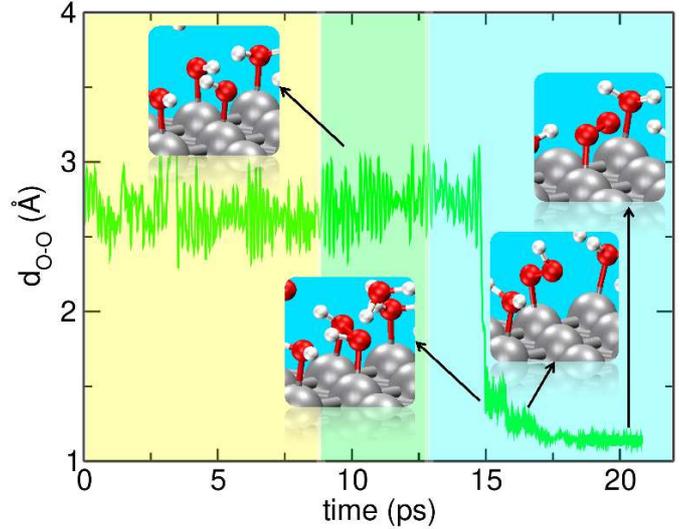}
\caption{Time evolution of the distance between the two O atoms involved in the OER. Representative snapshots of the 
intermediate adsorbates are provided in the insets. The background colors indicate the three potential regimes.}
\label{Fig4}
\end{center}
\end{figure}

%our OER reaction path
Next, we induce the OER by further increasing the bias potential and focus on the reaction mechanism.
As shown in Fig.\ \ref{Fig1}, more electronic charge is extracted from the cell and the system is driven 
out of the equilibrium at $Q_{\rm tot}$=0.62 $e^-$/Pt reaching a new state at
$Q_{\rm tot}$=1.04 $e^-$/Pt. During this transition, we observe the OER starting on surface 2 at 14.3 ps 
and leading to the formation of an adsorbed O$_2$ molecule within a few picoseconds. The OER is found 
to proceed through the following steps. At first, two hydroxyl ions are adsorbed at adjacent Pt active sites. 
Figure \ref{Fig4} shows the time evolution of the distance $d_{\rm O-O}$ between
the oxygen atoms of the adsorbed hydroxyl ions. The distance between the two ${\rm OH}_{\rm ads}$ fluctuates around 
2.7 {\AA}, close to the Pt-Pt distance of 2.83 {\AA}. Second, $d_{\rm O-O}$ shortens to 1.42 {\AA} upon 
the formation of a ${\rm H_2O_2}_{\rm ads}$ molecule.
Third, ${\rm H_2O_2}_{\rm ads}$ releases one hydrogen and transforms into ${\rm OOH}_{\rm ads}$, with $d_{\rm O-O}\cong 1.25$ {\AA}. 
Finally, the second hydrogen is released and the resulting ${\rm O}_{2\rm ads}$ shows a $d_{\rm O-O}$ distance of 1.13 {\AA}.
This distance is very close to the equilibrium distance of 1.20 {\AA} of gaseous O$_2$. Snapshots of the successive 
intermediate adsorbates are given in the insets of Fig.\ \ref{Fig4}.

During the dynamics, we simultaneously observe the first steps of the associative mechanism up to the 
formation of ${\rm O}_{\rm ads}$, but the subsequent steps of this mechanism do not occur on the time scale of our simulation. 
From the observed evolution, we thus draw the following mechanism for the OER on Pt:
\begin{align*}
^{*} + {\rm H_2O}              &\rightarrow {\rm OH}_{\rm ads} + ({\rm H}^{+} + e^-), \\
{\rm OH}_{\rm ads} + ^{*} + {\rm H_2O}           &\rightarrow 2{\rm OH}_{\rm ads} + ({\rm H}^{+} + e^-),  \\
2{\rm OH}_{\rm ads} \rightarrow {\rm H_2O_2}_{\rm ads} &\rightarrow {\rm OOH}_{\rm ads} + ({\rm H}^{+} + e^-),  \\
{\rm OOH}_{\rm ads}            &\rightarrow {\rm O}_{2\rm g} + ({\rm H}^{+} + e^-), 
\end{align*}
where the second step involves the adsorption of the second ${\rm OH}_{\rm ads}$ in the immediate vicinity of the 
${\rm OH}_{\rm ads}$ adsorbed in the first step.
The formation of the ${\rm H_2O_2}_{\rm ads}$ molecule occurs as an intermediate in the third step and is facilitated by the irregular 
${\rm H_2O}_{\rm ads}$/${\rm OH}_{\rm ads}$ hexagonal ML at the Pt surface, which allows for adjacent ${\rm OH}_{\rm ads}$. 
We remark that the present mechanism does not rely on the formation of an ${\rm O}_{\rm ads}$ intermediate. 
At variance, our work provides evidence for the occurrence of the hydrogen peroxide intermediate, as suggested by 
Wroblowa et al.\cite{wroblowa1976electroreduction} and supports the experimental observation of this species during the OER % 
\cite{gomez2012interaction,zhang2005controlling,stamenkovic2001structure,schmidt2001oxygen,schneider2008transport,seidel2009mesoscopic,markovic2002surface,markovic1999oxygen,grgur1997temperature}.

\begin{figure}[!t]
\begin{center}
\includegraphics*[width=1.0\linewidth, keepaspectratio=true]{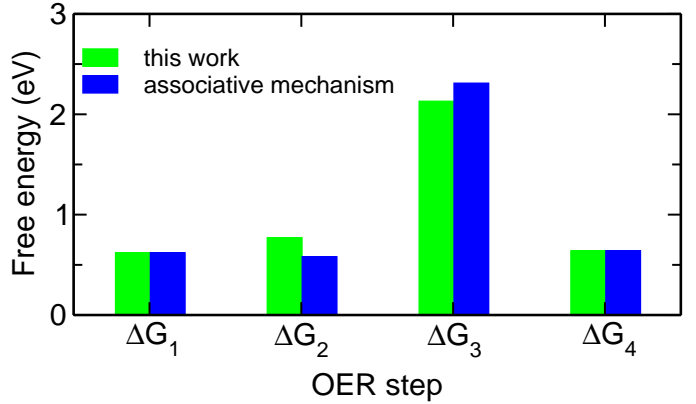}
\caption{Gibbs free energy steps corresponding to the intermediates of the OER on Pt(111).}
\label{Fig5}
\end{center}
\end{figure}

%Free energy calculations
The pathway revealed by the constant Fermi level molecular dynamics has been enforced by driving the potential 
to values for which the reaction could evolve spontaneously. Furthermore, our observation hinges on 
a single molecular dynamics trajectory. To corroborate our finding, it is therefore necessary to 
critically examine the identified reaction path by resorting to a method corresponding to the state of the art.
We are not aware of any generally accepted scheme to date involving bias potentials and explicit water that produces 
reliable free energies. We therefore resort to the widely used computational hydrogen electrode scheme \cite{norskov2004origin}.
Within this scheme, we calculate the Gibbs free energies, $\Delta G_1$ to $\Delta G_4$, corresponding to the four proton-coupled 
electron transfers.
The calculations are performed at the bare Pt(111) and the systems with the intermediate adsorbates are fully relaxed 
at the rVV10 level of theory. To enable a meaningful comparison of the calculated free energies with those of the associative 
mechanism in the literature \cite{norskov2004origin,gono2018surface}, we here use the CP2K code \cite{vandevondele2005quickstep},
in which the orbitals are described with an atom-centered Gaussian-type basis set and the electron density is expanded on an auxiliary 
plane-wave basis set.  We use triple-$\zeta$ MOLOPT basis sets \cite{vandevondele2007gaussian} for all elements.
The plane-wave energy cutoff is set to 800 Ry and the Brillouin zone is sampled at the $\Gamma$ point.
Goedecker-Teter-Hutter pseudopotentials are used to describe core-valence interactions \cite{goedecker1996separable}.
A detailed description of the adopted formulation is provided in the Supporting Information of Ref.\ \citenum{gono2018surface}.

The calculated free energies are illustrated in Fig.\ \ref{Fig5}. The presently identified mechanism differs from the associative 
one only in the second and third steps because of the different intermediates involved. 
In particular, the associative mechanism involves ${\rm O}_{\rm ads}$, which interacts strongly with the Pt electrode 
leading to a high value of $\Delta G_3$ \cite{norskov2004origin,tripkovic2010oxygen}. 
In the presently identified pathway, the OER overpotential of the overall reaction is still 
determined by the third proton-coupled electron transfer ($\Delta G_3$), but it is found to be lower by 0.18 eV. 
This energy gain results from the extra cost of adsorbing a second ${\rm OH}_{\rm ads}$ in the vicinity of a previously
adsorbed ${\rm OH}_{\rm ads}$, which leads to a corresponding increase in $\Delta G_2$.
Hence, these results indicate that the adopted simulation scheme has revealed a viable reaction pathway with 
competitive Gibbs free energy steps. 

%Conclusions
In summary, constant Fermi level molecular dynamics at the Pt(111)/metal interface reveal  
that adsorbates arrange in a honeycomb lattice at the Pt surface with an ${\rm OH}_{\rm ads}$ coverage of 1/3 ML. 
We find that the ${\rm H_2O}_{\rm ads}$ and ${\rm OH}_{\rm ads}$ do not respect the perfect alternation at the Pt surface
leading to the adsorption of ${\rm OH}_{\rm ads}$ at adjacent Pt sites. Consequently, this opens the way to the formation of 
an adsorbed hydrogen peroxide molecule as an intermediate species, which facilitates the OER.
Free energy calculations indicate that the pathway involving a hydrogen peroxide shows 
a reaction overpotential that is lower by 0.2 eV compared to the associative mechanism. 
These results demonstrate the powerfulness of constant Fermi level molecular dynamics as un unbiased 
tool for identifying low-energy pathways of catalytic electrochemical reactions.

%\textcolor{red}{In addition, these results provides a possible hypothesis about ${\rm H_2O_2}$ formation during 
%water oxidation which might be due to the formation of imperfect ${\rm OH}_{\rm ads}$ and ${\rm H_2O}_{\rm ads}$ 
%alternation at the Pt surface.}

This project has received funding from the European Union's Seventh Framework Programme 
for research, technological development and demonstration under grant agreement no.\ 291771 (call 2014). 
This work has been performed in the context of the National Center
of Competence in Research (NCCR) ``Materials' Revolution:
Computational Design and Discovery of Novel Materials (MARVEL)'' of the Swiss National 
Science Foundation. We used computational resources of CSCS, SCITAS, and CSEA-EPFL.

%\bibliography{bib-paper}

\begin{thebibliography}{10}
\expandafter\ifx\csname url\endcsname\relax
  \def\url#1{\texttt{#1}}\fi
\expandafter\ifx\csname urlprefix\endcsname\relax\def\urlprefix{URL }\fi
\expandafter\ifx\csname href\endcsname\relax
  \def\href#1#2{#2} \def\path#1{#1}\fi

\bibitem{fontaras2007quantitative}
G.~Fontaras, Z.~Samaras, A quantitative analysis of the european automakers
  voluntary commitment to reduce co$_2$ emissions from new passenger cars based
  on independent experimental data, Energy Pol. 35~(4) (2007) 2239--2248.

\bibitem{basha2009review}
S.~A. Basha, K.~R. Gopal, S.~Jebaraj, A review on biodiesel production,
  combustion, emissions and performance, Renew. Sustain. Energy Rev. 13~(6-7)
  (2009) 1628--1634.

\bibitem{samanta2011post}
A.~Samanta, A.~Zhao, G.~K. Shimizu, P.~Sarkar, R.~Gupta, Post-combustion co$_2$
  capture using solid sorbents: a review, Ind. Eng. Chem. Res. 51~(4) (2011)
  1438--1463.

\bibitem{kinoshita1992electrochemical}
K.~Kinoshita, Electrochemical oxygen technology, Vol.~30, John Wiley \& Sons,
  1992.

\bibitem{norskov2004origin}
J.~K. N{\o}rskov, J.~Rossmeisl, A.~Logadottir, L.~Lindqvist, J.~R. Kitchin,
  T.~Bligaard, H.~Jonsson, Origin of the overpotential for oxygen reduction at
  a fuel-cell cathode, J. Phys. Chem. B 108~(46) (2004) 17886--17892.

\bibitem{li2013ph}
M.~F. Li, L.~W. Liao, D.~F. Yuan, D.~Mei, Y.-X. Chen, ph effect on oxygen
  reduction reaction at pt (1 1 1) electrode, Electrochim. Acta 110 (2013)
  780--789.

\bibitem{roques2005potential}
J.~Roques, A.~B. Anderson, V.~S. Murthi, S.~Mukerjee, Potential shift for
  oh(ads) formation on the pt skin on pt$_3$co(111) electrodes in acid theory
  and experiment, J. Electrochem. Soc. 152~(6) (2005) E193--E199.

\bibitem{bondarenko2011pt}
A.~S. Bondarenko, I.~E. Stephens, H.~A. Hansen, F.~J. P{\'e}rez-Alonso,
  V.~Tripkovic, T.~P. Johansson, J.~Rossmeisl, J.~K. N{\o}rskov,
  I.~Chorkendorff, The pt(111)/electrolyte interface under oxygen reduction
  reaction conditions: an electrochemical impedance spectroscopy study,
  Langmuir 27~(5) (2011) 2058--2066.

\bibitem{karlberg2005interaction}
G.~Karlberg, G.~Wahnstr{\"o}m, An interaction model for oh$^+$ h$_2$o-mixed and
  pure h$_2$o overlayers adsorbed on pt(111), J. Chem. Phys. 122~(19) (2005)
  194705.

\bibitem{liu2016mechanism}
S.~Liu, M.~G. White, P.~Liu, Mechanism of oxygen reduction reaction on pt(111)
  in alkaline solution: Importance of chemisorbed water on surface, J. Chem.
  Phys. C 120~(28) (2016) 15288--15298.

\bibitem{wroblowa1976electroreduction}
H.~S. Wroblowa, G.~Razumney, et~al., Electroreduction of oxygen: a new
  mechanistic criterion, J. Electroanal. Chem. Interf. Elec. 69~(2) (1976)
  195--201.

\bibitem{gomez2012interaction}
A.~G{\'o}mez-Mar{\'\i}n, K.~Schouten, M.~Koper, J.~Feliu, Interaction of
  hydrogen peroxide with a pt(111) electrode, Electrochem. Commun. 22 (2012)
  153--156.

\bibitem{zhang2005controlling}
J.~Zhang, M.~B. Vukmirovic, Y.~Xu, M.~Mavrikakis, R.~R. Adzic, Controlling the
  catalytic activity of platinum-monolayer electrocatalysts for oxygen
  reduction with different substrates, Angew. Chem. 117~(14) (2005) 2170--2173.

\bibitem{stamenkovic2001structure}
V.~Stamenkovic, N.~Markovic, P.~Ross~Jr, Structure-relationships in
  electrocatalysis: oxygen reduction and hydrogen oxidation reactions on
  pt(111) and pt(100) in solutions containing chloride ions, J. Electroanal.
  Chem. 500~(1-2) (2001) 44--51.

\bibitem{schmidt2001oxygen}
T.~Schmidt, U.~Paulus, H.~A. Gasteiger, R.~Behm, The oxygen reduction reaction
  on a pt/carbon fuel cell catalyst in the presence of chloride anions, J.
  Electroanal. Chem. 508~(1-2) (2001) 41--47.

\bibitem{schneider2008transport}
A.~Schneider, L.~Colmenares, Y.~Seidel, Z.~Jusys, B.~Wickman, B.~Kasemo,
  R.~Behm, Transport effects in the oxygen reduction reaction on
  nanostructured, planar glassy carbon supported pt/gc model electrodes, Phys.
  Chem. Chem. Phys. 10~(14) (2008) 1931--1943.

\bibitem{seidel2009mesoscopic}
Y.~Seidel, A.~Schneider, Z.~Jusys, B.~Wickman, B.~Kasemo, R.~Behm, Mesoscopic
  mass transport effects in electrocatalytic processes, Faraday Discuss. 140
  (2009) 167--184.

\bibitem{markovic2002surface}
N.~Markovi{\'c}, P.~Ross~Jr, Surface science studies of model fuel cell
  electrocatalysts, Surf. Sci. Rep. 45~(4-6) (2002) 117--229.

\bibitem{markovic1999oxygen}
N.~Markovi{\'c}, H.~Gasteiger, B.~Grgur, P.~Ross, Oxygen reduction reaction on
  pt(111): effects of bromide, J. Electroanal. Chem. 467~(1-2) (1999) 157--163.

\bibitem{grgur1997temperature}
B.~Grgur, N.~Markovi{\'c}, P.~Ross, Temperature-dependent oxygen
  electrochemistry on platinum low-index single crystal surfaces in acid
  solutions, Can. J. Chem. 75~(11) (1997) 1465--1471.

\bibitem{wang2004kinetic}
J.~Wang, N.~Markovic, R.~Adzic, Kinetic analysis of oxygen reduction on pt(111)
  in acid solutions: intrinsic kinetic parameters and anion adsorption effects,
  J. Phys. Chem. B 108~(13) (2004) 4127--4133.

\bibitem{tripkovic2010oxygen}
V.~Tripkovi{\'c}, E.~Sk{\'u}lason, S.~Siahrostami, J.~K. N{\o}rskov,
  J.~Rossmeisl, The oxygen reduction reaction mechanism on pt (1 1 1) from
  density functional theory calculations, Electrochim. Acta 55~(27) (2010)
  7975--7981.

\bibitem{tripkovic2011standard}
V.~Tripkovic, M.~E. Bj{\"o}rketun, E.~Sk{\'u}lason, J.~Rossmeisl, Standard
  hydrogen electrode and potential of zero charge in density functional
  calculations, Phys. Rev. B 84~(11) (2011) 115452.

\bibitem{skulason2007density}
E.~Sk{\'u}lason, G.~S. Karlberg, J.~Rossmeisl, T.~Bligaard, J.~Greeley,
  H.~J{\'o}nsson, J.~K. N{\o}rskov, Density functional theory calculations for
  the hydrogen evolution reaction in an electrochemical double layer on the
  pt(111) electrode, Phys. Chem. Chem. Phys. 9~(25) (2007) 3241--3250.

\bibitem{rossmeisl2013ph}
J.~Rossmeisl, K.~Chan, R.~Ahmed, V.~Tripkovi{\'c}, M.~E. Bj{\"o}rketun, ph in
  atomic scale simulations of electrochemical interfaces, Phys. Chem. Chem.
  Phys. 15~(25) (2013) 10321--10325.

\bibitem{bouzidjctc2017}
A.~Bouzid, A.~Pasquarello, Redox levels through constant fermi-level ab initio
  molecular dynamics, J. Chem. Theory Comput. 13~(4) (2017) 1769--1777.

\bibitem{bouzid2018atomic}
A.~Bouzid, A.~Pasquarello, Atomic-scale simulation of electrochemical processes
  at electrode/water interfaces under referenced bias potential, J. Phys. Chem.
  Lett. 9~(8) (2018) 1880--1884.

\bibitem{otani2006first}
M.~Otani, O.~Sugino, First-principles calculations of charged surfaces and
  interfaces: A plane-wave nonrepeated slab approach, Phys. Rev. B 73~(11)
  (2006) 115407.

\bibitem{otani2008structure}
M.~Otani, I.~Hamada, O.~Sugino, Y.~Morikawa, Y.~Okamoto, T.~Ikeshoji, Structure
  of the water/platinum interface -- a first principles simulation under bias
  potential, Phys. Chem. Chem. Phys. 10~(25) (2008) 3609--3612.

\bibitem{otani2008electrode}
M.~Otani, I.~Hamada, O.~Sugino, Y.~Morikawa, Y.~Okamoto, T.~Ikeshoji, Electrode
  dynamics from first principles, J. Phys. Soc. Jpn. 77~(2) (2008) 024802.

\bibitem{ikeshoji2017first}
T.~Ikeshoji, T.~Uchida, M.~Otani, M.~Osawa, First-principles molecular dynamics
  simulation for electrochemical hydrogen production by 4,4-bipyridine
  molecular catalyst on silver electrode, J. Electroanal. Chem. 800 (2017)
  13--18.

\bibitem{bonnet2012first}
N.~Bonnet, T.~Morishita, O.~Sugino, M.~Otani, First-principles molecular
  dynamics at a constant electrode potential, Phys. Rev. Lett. 109~(26) (2012)
  266101.

\bibitem{bouzid2017electron}
A.~Bouzid, A.~Pasquarello, Electron trap states at ingaas/oxide interfaces
  under inversion through constant fermi-level ab initio molecular dynamics, J.
  Phys. Condens. Matter 29~(50) (2017) 505702.

\bibitem{bouzid2017identification}
A.~Bouzid, A.~Pasquarello, Identification of semiconductor defects through
  constant-fermi-level ab initio molecular dynamics: Application to gaas, Phys.
  Rev. Appl. 8 (2017) 014010.

\bibitem{vydrov2010nonlocal}
O.~A. Vydrov, T.~Van~Voorhis, Nonlocal van der waals density functional: The
  simpler the better, J. Chem. Phys. 133~(24) (2010) 244103.

\bibitem{sabatini2013nonlocal}
R.~Sabatini, T.~Gorni, S.~de~Gironcoli, Nonlocal van der waals density
  functional made simple and efficient, Phys. Rev. B 87~(4) (2013) 041108.

\bibitem{miceli2015isobaric}
G.~Miceli, S.~de~Gironcoli, A.~Pasquarello, Isobaric first-principles molecular
  dynamics of liquid water with nonlocal van der waals interactions, J. Chem.
  Phys. 142~(3) (2015) 034501.

\bibitem{TM}
N.~{Troullier}, J.~L. {Martins}, Efficient pseudopotentials for plane-wave
  calculations, Phys. Rev. B 43~(3) (1991) 1993--2006.
\newblock \href {http://dx.doi.org/10.1103/PhysRevB.43.1993}
  {\path{doi:10.1103/PhysRevB.43.1993}}.

\bibitem{giannozzi2009quantum}
P.~Giannozzi, S.~Baroni, N.~Bonini, M.~Calandra, R.~Car, C.~Cavazzoni,
  D.~Ceresoli, G.~L. Chiarotti, M.~Cococcioni, I.~Dabo, et~al., Quantum
  espresso: a modular and open-source software project for quantum simulations
  of materials, J. Phys. Condens. Matter 21~(39) (2009) 395502.

\bibitem{taylor2006first}
C.~D. Taylor, S.~A. Wasileski, J.-S. Filhol, M.~Neurock, First principles
  reaction modeling of the electrochemical interface: Consideration and
  calculation of a tunable surface potential from atomic and electronic
  structure, Phy. Rev. B 73~(16) (2006) 165402.

\bibitem{hansen2008surface}
H.~A. Hansen, J.~Rossmeisl, J.~K. N{\o}rskov, Surface pourbaix diagrams and
  oxygen reduction activity of pt, ag and ni(111) surfaces studied by dft,
  Phys. Chem. Chem. Phys. 10~(25) (2008) 3722--3730.

\bibitem{gono2018surface}
P.~Gono, J.~Wiktor, F.~Ambrosio, A.~Pasquarello, Surface polarons reducing
  overpotentials in the oxygen evolution reaction, ACS Catal. 8 (2018)
  5847--5851.

\bibitem{vandevondele2005quickstep}
J.~VandeVondele, M.~Krack, F.~Mohamed, M.~Parrinello, T.~Chassaing, J.~Hutter,
  Quickstep: Fast and accurate density functional calculations using a mixed
  gaussian and plane waves approach, Comput. Phys. Commun. 167~(2) (2005)
  103--128.

\bibitem{vandevondele2007gaussian}
J.~VandeVondele, J.~Hutter, Gaussian basis sets for accurate calculations on
  molecular systems in gas and condensed phases, J. Chem. Phys. 127~(11) (2007)
  114105.

\bibitem{goedecker1996separable}
S.~Goedecker, M.~Teter, J.~Hutter, Separable dual-space gaussian
  pseudopotentials, Phys. Rev. B 54~(3) (1996) 1703.

\end{thebibliography}
%\bibliographystyle{elsarticle-num}

\end{document}